\documentclass[showpacs,12pt,amsmath,amssymb,aps,onecolumn]{revtex4-1}
\usepackage{graphics}
\usepackage{graphicx}
\usepackage{subfigure}
\usepackage{txfonts}
\usepackage{bm}   
\usepackage{amsmath,mathrsfs}
\usepackage{algorithm2e}

\usepackage{amssymb}
\usepackage{mathrsfs}
\usepackage{color}

\usepackage{multirow}
\begin{document}

\title{Testing Violation of the Leggett-Garg-type inequality in neutrino oscillations of Daya-Bay experiment\thanks{This work is partly supported by the Key Research Program of Frontier Sciences, CAS, under the Grants Number NO. QYZDY-SSW-SLH006 of Chinese Academy of Sciences.}}

\author{Qiang Fu$^{1,2,3}$}
\author{Xurong Chen$^{1}$}\email{xchen@impcas.ac.cn}
\affiliation{
$^1$ Institute of Modern Physics, Chinese Academy of Sciences, Lanzhou, 730000, China\\
$^2$ Lanzhou University, Lanzhou 730000, China\\
$^3$ University of Chinese Academy of Sciences, Beijing 100049, China \\
}
\date{\today}

\begin{abstract}
The Leggett-Garg inequality (LGI), derived under the assumption of realism, acts as the temporal Bell's inequality. It is studied in electromagnetic and strong interaction like photonics, superconducting qu-bits and nuclear spin. Until the weak interaction two-state oscillations of neutrinos affirmed the violation of Leggett-Garg-type inequalities (LGtI). We make an empirical test for the deviation of experimental results with the classical limits by analyzing the survival probability data of reactor neutrinos at a distinct range of baseline dividing energies, as an analog to a single neutrino detected at different time. A study of the updated data of Daya-Bay experiment unambiguously depicts an obvious cluster of data over the classical bound of LGtI and shows a $6.1\sigma$ significance of the violation of them.
\end{abstract}

\maketitle 
To be published on European Physical Journal C.
\section{Introduction}

\label{SecI}
Nonclassical features of the quantum system have experienced extensive study since the inception of quantum mechanics. After a long debate between the believer of local realism and quantum mechanics, a breakthrough study, Bell's inequality (BI), was provided by Bell~\cite{Bell:1964kc}. The unique feature of BI is its testable formula from the consequence of the famous hypothesis called the local realism (LR). The LR believers assume that any observable value of an object, even not detected, must have a definite value and that results of any individual measurement of the observables remain unaffected if they were in space-like separation. Extensive experimental investigations~\cite{Hensen:2015ccp,Weihs:1998gy,Aspect:1981nv} over past several decades tested the violation of BI. These studies conclude that any local realism view of a microscopic object needs to be nonlocal. Based on these studies of BI, Leggett and Garg furtherly derived a new series of inequalities~\cite{Leggett:1985zz} on the assumption of Macrorealism (MR), now known as the Leggett–Garg inequalities (LGIs), that any system behaving as a macroscopic realism must obey. From the structure of the LGIs, we can see them as a analogue of Bell's inequalities in temporal notion, which also makes it possible to implement a rigorous test of quantum mechanics in a macroscopic level which is usually very difficult in designing experiments in space-like separation condition. By testing the LGtI, we can also perform a rigorously loophole-free test of quantum mechanics~\cite{Hensen:2015ccp,PhysRevLett.115.250402,PhysRevLett.115.250401,Gallicchio:2013iva}. 

Besides the nonlocal behavior and quantum correlation between different particles, single particle states can also exist entanglement by the flavor transition\cite{Blasone:2007vw}. Since the oscillation of neutrino flavors, it offers us an ideal sources to test the quantum mechanics in weak interaction and under a macroscopic view. For two-flavor neutrino oscillation, a two level state's matrix can be expressed in the form
\begin{eqnarray}
\begin{aligned}
\rho = \frac{1}{4}\left[I\otimes I + (\rm{r}\cdot \rm{\sigma})\otimes I + I\otimes(\rm{s}\cdot \rm{\sigma}) +\sum_{n,m=1}^{3}T_{mn}(\sigma_{m}\otimes\sigma_{n})\right].
\label{eq:1_1}
\end{aligned}
\end{eqnarray}
Where the elements of matrix T are $T_{mn} = Tr\left[\rho(\delta_{m}\otimes \delta_{n})\right]$. For two qu-bit situation, many kinds of quantum correlations~\cite{PhysRevLett.78.5022,PhysRevLett.80.2245,PhysRevLett.105.190502} like the entanglement, fidelity, quantum discord and geometric discord have been derived to get their general expressions. Mermin and Svetlichny~\cite{PhysRevLett.65.1838,PhysRevD.35.3066} inequalities were derived for three or even many body system that have two macroscopic distinct states. Using these results of two level states, a series quantum-information theory calculation \cite{PhysRevA.88.022115,Banerjee2015,Alok:2014gya,Gangopadhyay2017} has been applied on neutrino flavor oscillation in last several years. These quantum correlations can be directly linked to the probabilities of flavor oscillation, which lead to violation of classical boundary limits when assuming the neutrino oscillation mixing angle is not vanishing. 

Neutrino flavor oscillation is such a special process that it merely only affected by their own properties like mass square differences, mixing angles and the energies. Since neutrinos just interact with matters by weak interaction with quite low cross-section. The influence of the environment act on neutrinos' propagation is much more negligible comparing with optical or electrical system, which makes neutrinos an ideal particle on testing the LGIs. As the mass eigenstate of a neutrino is not the same with its flavor eigenstate, during propagation, neutrinos undergo flavor mixing among the three flavored eigenstates. The MINOS experiment has been studied in Ref.~\cite{PhysRevLett.117.050402}, which observed the violation of Leggett-Garg-type inequalities, K3 and K4 terms, in a significance greater than $6\sigma$~\cite{PhysRevLett.117.050402}. The MINOS experiment is an accelerator neutrino experiment using decay in flight neutrinos with fixed baseline distance 735 km and a large range of $\nu_{\mu}$ energy from 0.5 GeV to 50 GeV, which happens to cover the largest violation of LGIs K3 and K4. The Daya-Bay Collaboration reported an updated data analysis of electron anti-neutrino disappearance channel~\cite{PhysRevLett.115.111802}, which gives a best fit of $sin^{2}2\theta_{13}=0.084\pm 0.005$, we will investigate whether the Daya-Bay Reactor Neutrino Experiment can observe the violation of LGIs.
\section{The Leggett-Garg-type inequalities}
\label{SecII}
We focus on the simplest L-G-type inequality which is constructed as follows. Consider a system with two absolutely distinguishable states correspond to an observable quantity $Q(t)$ can be two different value +1 or -1. Assuming that whenever the system was being measured, the observable quantity occupies a value of either +1 or -1 for being in state 1 or 2 respectively. Then we can define a macroscopic observable $Q(t)$ for the macroscopic system. And $C_{t_i,t_j}=\left< Q(t_i)Q(t_j)\right>$ as its two-time correlation function, where $Q(t_i)$ and $Q(t_j)$ are observable quantity's value when being measured at time $t_i$ and time $t_j$. In this work we consider the two state as the survival of electron anti-neutrino and the disappearance of the electron anti-neutrino. When the neutrinos being created in the reactor by beta decay process, they are totally in the state of the flavor eigenstate. Since the PMNS matrix does not change with time, the two-flavor neutrino oscillation obey the same survival probability. We shall introduce this stationary assumption~\cite{PhysRevA.52.R2497} which requires the evolution of the neutrino for different ordered time intervals are the same. And then the $C(t_i,t_j) = C(t_i-t_j)$ (if $t_i < t_j$). Next, consider a sequence of times $t_1, t_2, t_3$ and $t_4$(here, $t_1<t_2<t_3<t_4$). If we take a series of measurements for $Q(t)$ in these four times, it is straightforward to determine four time correlations ($C_{12}, C_{23}, C_{34}$ and $C_{14}$). Then it is possible to adopt the stationary condition on standard LGI procedure leading to K4 LG-type inequality involving four correlation functions. For any sequence of measurements, any $Q(t_i)$ has the definite observable value, regardless of the choice of the pair $Q(t_i)Q(t_j)$ it belongs to. So, combination $Q(t_1)Q(t_2)+Q(t_2)Q(t_3)+Q(t_3)Q(t_4)-Q(t_1)Q(t_4)$ lies always between -2 and +2. Similarly, $K_3$ inequality lies between -1 and +1. If all the terms in the above formula are replaced by time correlations (average), the Leggett-Garg-type inequalities are in form
\begin{eqnarray}
\begin{aligned}
&K_3 \equiv C_{12}+C_{23}-C_{13} \leq 1. \\
&K_4 \equiv C_{12}+C_{23}+C_{34}-C_{14} \leq 2. 
\label{eq:2_1}
\end{aligned}
\end{eqnarray}
The above inequalities imposes a constraint on macroscopic realism on temporal separated joint probabilities in any two-state system.
\section{Oscillating neutrinos}
\label{SecIII}
\begin{figure}[htp]
\begin{center}
\includegraphics[width=0.55\textwidth]{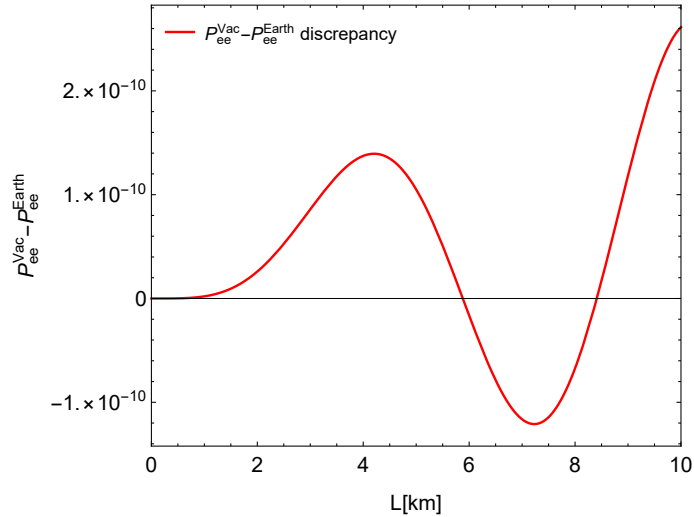}
\caption{(color online). Illustration of the matter effect of the Earth in the interval of 0~10 km. The plot is the discrepancy of vacuum $P_{ee}^{Vac}$ and the Earth $P_{ee}^{Earth}$. The neutrino is 6 MeV monoenergetic, and the $\rho(x) = 2.65~g/cm^{3}$ for the standard rock.}
\label{Fig:3_0}
\end{center}
\end{figure}

It is extensively verified that the flavor component of a neutrino oscillates during its propagation. The oscillation properties of different neutrino flavors are determined by there mixing angles ($\theta_{12}, \theta_{23}$, and $\theta_{13}$), a CP phase of the Pontecorvo-Maki-Nakagawa-Sakata matrix and there mass-squared differences ($\Delta m_{32}^2, \Delta m_{21}^2$)~\cite{Pontecorvo:1967fh, Maki:1962mu}. Here, we will treat the Leggett-Garg-type inequalities using the updated measurement of $\overline{\nu_e}$ survival channel in the results of the Daya-Bay experiment, where give the newest best fit of mixing angle $\theta_{13}$ with large significance~\cite{PhysRevLett.115.111802}. The measurement was updated later with full detector configuration~\cite{PhysRevLett.115.111802}. For $\theta_{13}$ measurement, they used baseline length dividing $\overline{\nu_e}$ energy as the variable to depict the survival probability of $\overline{\nu_e}$ as 
\begin{eqnarray}
\begin{aligned}
P_{\overline{\nu}_e \rightarrow \overline{\nu}_e} = 1-cos^{4}\theta_{13}sin^{2}2\theta_{12}sin^{2}\frac{1.267\Delta m_{21}^{2}L}{E}-sin^{2}2\theta_{13}sin^{2}\frac{1.267\Delta m_{ee}^{2}L}{E}.
\label{eq:3_1}
\end{aligned} 
\end{eqnarray}
where $E$ is the energy of $\overline{\nu_e}$ in MeV, L is the propagation distance between near and far point detector, $\theta_{12}$ is the solar neutrino mixing angle and $\Delta m_{21}^{2}$ is their mass-squared difference in $eV^2$. Notice that $\Delta m_{ee}^{2}$ is an effective mass-squared difference~\cite{PhysRevD.72.013009} in electron anti-neutrino disappearance with the form of
\begin{eqnarray}
\begin{aligned}
\Delta m_{ee}^{2} = cos^2\theta_{12}\Delta m_{31}^2+sin^2\theta_{12}\Delta m_{32}^2.
\label{eq:3_2}
\end{aligned} 
\end{eqnarray}
Since the $m_{12}^2 = (7.50\pm 0.20)\times 10^{-5}~\rm{eV^2}$, while the $m_{ee}^2 = (2.42\pm 0.11)\times 10^{-3}~\rm{eV^2}$ according to Ref.~\cite{PhysRevLett.115.111802}, we can choose a appropriate value of the ratio L/E to make one of a term of $sin^{2}\frac{1.267\Delta m^{2}L}{E}$ vanishing. For Daya-Bay experiment, the effect of parameter $\theta_{12}$ becomes far less enough than that from $\theta_{13}$ to make $\theta_{12}$ regarded as negligibly small. Given that there is an initial pure electron anti-neutrino source, after a time t propagation, the survival probability of $\overline{\nu}_e$ will be:
\begin{eqnarray}
\begin{aligned}
P_{\overline{\nu}_e \rightarrow \overline{\nu}_e} = 1-sin^{2}2\theta_{13}sin^{2}\frac{1.267\Delta m_{ee}^{2}ct}{E}.
\label{eq:3_3}
\end{aligned} 
\end{eqnarray}
Although there is the MSW (Mikheyev-Smirnov-Wolfenstein) effect (usually called matter effect) during the propagation of neutrino in matter. The effect is only significant for high energy neutrinos and long range of matter, like the solar neutrino experiment. The KamLAND and Super-K's $P_{ee}$ day-night discrepancy are only obvious for larger than 6 MeV neutrinos\cite{PhysRevD.56.1792,Vissani:2017dto}. Further more, the solar neutrino experiments involve the matter effect caused by the electron in the solar, which electron density $\epsilon_{\odot}$ is great larger than that in the Earth. Generally speaking, a neutrino vector of state in flavor basis $\left|\nu(t)\right> = (\nu_{e}(t)~\nu_{\mu}(t)~\nu_{\tau}(t))^{T}$ flows the $\rm{Schr\ddot{o}dinger}$ equation:
\begin{eqnarray}
\begin{aligned}
i\frac{d}{dt}\left|\nu(t)\right> = \mathcal{H} \left|\nu(t)\right>
\label{eq:3_4}
\end{aligned}
\end{eqnarray} 
where the Hamiltonian can be replaced by an effective one as
\begin{eqnarray}
\begin{aligned}
\mathcal{H} \simeq \frac{1}{2E}\rm{U}~\rm{diag}(0, \Delta m_{21}^2, \Delta m_{31}^2)U^{\dagger} + \rm{diag}(V,0,0).
\label{eq:3_5}
\end{aligned}
\end{eqnarray} 
where $V$ is the effective charged potential contribution to $\nu_e$~\cite{PhysRevD.17.2369} given in the form
\begin{eqnarray}
\begin{aligned}
V(x) \simeq 7.56\times 10^{-14}\left(\frac{\rho(x)}{\rm{g/cm^{3}}}\right)Y_{e}(x)~\rm{eV},
\label{eq:3_6}
\end{aligned}
\end{eqnarray} 
where $\rho(x)$ is the matter density along the track path of the neutrino,$Y_{e}(x)$ (for the Earth $\simeq$ 0.5) is the number of electrons normalized to the number of nucleons. For the matter of constant density, the series expansion for three-flavor neutrino oscillation probabilities can be derived from the Hamiltonian~Eq.(\ref{eq:3_5})~\cite{Akhmedov:2004ny}. For $\nu_{e}$ survival, the survival probability expansion to second order is
\begin{eqnarray}
\begin{aligned}
P_{ee} = 1- \alpha sin^{2}2\theta_{12}\frac{sin^{2}A\Delta}{A^{2}} - 4sin^{2}\theta_{13}\frac{sin^{2}(A-1)\Delta}{(A-1)^{2}},
\label{eq:3_7}
\end{aligned}
\end{eqnarray} 
where $\alpha = \frac{\Delta m_{21}^{2}}{\Delta m_{31}^{2}} \simeq 0.0297$, and the abbreviation for $A$ and $\Delta$ is
\begin{eqnarray}
\begin{aligned}
&\Delta \equiv \frac{1.27\Delta m_{31}^{2}L}{E}~\rm{\frac{[ev^{2}][km]}{[GeV]}}, \\
&A \equiv \frac{2EV}{\Delta m_{31}^{2}\times 10^{-3}}~\rm{\frac{[GeV][eV]}{[eV^{2}]}}. 
\label{eq:3_8}
\end{aligned}
\end{eqnarray}
For this Daya-Bay analysis, we calculate the discrepancy of the $P_{ee}$ probability of a $6~MeV$ neutrino in the range of 0 to 10 km covering the range of the experiment, about 2 km. From Fig.~\ref{Fig:3_0}, we can draw a conclusion that the matter effect is too small to be included in ``short'' baseline.
 
According to the expansion of the $P_{ee}$ with matter effect mentioned above, we will use the oscillation probability just in the vacuum. Using the stationary assumption, one can derive the four joint probabilities $P_{\nu_{\alpha},\nu_{\beta}}(t_i,t_j)$, here $\alpha$ and $\beta$ stands for $\overline{\nu}_e$ and another neutrino flavor, i and j are from 1 to 4 defined above. The two-time correlation function $C_{12}$ in this $P_{\overline{\nu}_e \rightarrow \overline{\nu}_e}$ is given by
\begin{eqnarray}
\begin{aligned}
C_{12} = 1-\left[sin2\theta_{13}sin\left(\frac{1.267\Delta m_{ee}^{2}}{E}c(t_2-t_1)\right)\right]^2 = 2P_{\overline{\nu}_e \rightarrow \overline{\nu}_e}(t_2-t_1)-1.
\label{eq:3_9}
\end{aligned} 
\end{eqnarray}  
Similarly, the correlation functions $C_{23}$, $C_{34}$, and $C_{14}$ can be calculated in the same way. Using the Eq.~\ref{eq:3_3}, the quantity $K_{n}^Q$ can be evaluated as defined in Eq.~\ref{eq:2_1}. By choosing the time intervals in a particular way, we can achieve maximum value of $K_4$ when $t_4-t_3=t_3-t_2=t_2-t_1=\delta t$. Under this condition, the correlation functions depend on the baseline length $L$ and neutrino energy $E_{\nu}$. We select the neutrinos measured $L_{eff}/E$ to make the oscillation phase $\psi_a=\frac{1.267\Delta m^{2}}{E}c(\delta t)$ obey the sum rule: $\psi_{12}+\psi_{23}+\psi_{34}=\psi_{14}$. With an experimental arrangement in which measurements occur at some fixed distance from the neutrino sources. Assuming the neutrino begins in the pure $\left|\overline{\nu}_e\right>$
\begin{eqnarray}
\begin{aligned} 
K_{n}^Q = -2+2\sum_{a=1}^{n-1}P_{ee}(\psi_a)-2P_{ee}(\sum_{a=1}^{n-1}\psi_a).
\label{eq:3_10}
\end{aligned} 
\end{eqnarray}
Where the n can be 3 or 4 in this paper, which corresponds to $K_3$ or $K_4$ LGI. In quantum mechanical, the commutators of operators can be nonvanishing. While in a classical system, operators observable values must commute, then the macrorealism derived $K_n$ will become:
\begin{eqnarray}
\begin{aligned} 
K_{n}^C = \sum_{a=1}^{n-1}C_{i,i+1}-\prod_{a=1}^{n-1}C_{i,i+1}.
\label{eq:3_11}
\end{aligned} 
\end{eqnarray} 

The Daya-Bay Collaboration released updated oscillation results as a function of effective baseline distance $L_{eff}$ over the average energy $\left<E_{\nu}\right>$ in bins~\cite{PhysRevLett.115.111802}. For their six anti-neutrino detectors (ADs) placed in three separate experimental halls (EHs) and three nuclear reactors neutrino sources, the effective baseline varies for each detected anti-neutrino. The Daya-Bay experiment covers energy between 1 to 8 MeV. The range of effective baseline and energy correspond to a phase range of ~$\left(0,3/4\pi\right)$, within which the violations of LGI will be observed near the minimum point of the anti-neutrino survival probability.  

\begin{figure}[htp]
\begin{center}
\includegraphics[width=0.55\textwidth]{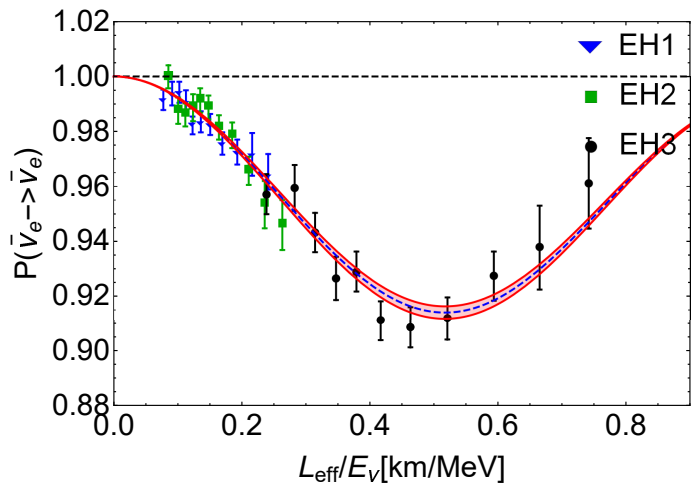}
\caption{(color online). The electron anti-neutrino survival probability versus effective propagation distance $L_{eff}$ over anti-neutrino energy $E_{\nu}$. The dashed blue curve indicates the prediction fitting directly to the measured Daya-Bay values of $P_{ee}$. The red band indicates a $1\sigma$ confidence interval around the fitted prediction. The blue triangles, green rectangles and black dots are the binned data of Daya-Bay EH1, EH2 and EH3 from Ref.~\cite{PhysRevLett.115.111802} respectively.}
\label{Fig:3_1}
\end{center}
\end{figure}

To test the violations of the $K_3$ and $K_4$ inequalities, we make an operation on the data from the Daya-Bay neutrino experiment. Daya-Bay experiment extracted the survival probabilities of neutrinos using Daya-Bay and Ling-Ao nuclear power stations' reactors. We use all the measurement positions including EH1, EH2 and EH3 of Daya-Bay. The reactors provide different sources of neutrinos with several fixed baseline and an energy spectrum with peaks. We make a $\theta_{13}$ fit over the Daya-Bay updated data and get the fit error band and center value of $P_{ee}$ shown in Fig.~\ref{Fig:3_1}. With the best fit of $sin^{2}2\theta_{13}$ and the $1\sigma$ error band of it, we generated a large sets of pseudodata. Then we select all sets of data points in Fig~\ref{eq:3_1} which obey the sum rule of phase with the precision of $0.5\%$ ($\psi_1+\psi_2 \in \psi_3$) and $0.1\%$ ($\psi_1+\psi_2+\psi_3 \in \psi_4$) for the $K_3$ and $K_4$ respectively. For $K_3$ ($K_4$) situation, 48 (56) correlation triples (quadruples) satisfy the sum rule. While the updated measurement only includes the static errors and we simply assumed that the errors at small phase of the oscillation probability are the fitting error.

The violation of Leggett-Garg-type inequalities has been tested and confirmed by the MINOS experiment with the $K_3$ and $K_4$ being inconsistent with realism prediction over $5\sigma$~\cite{PhysRevLett.117.050402}. Since the violation of Leggett-Garg-type inequalities happens when the mixing angle of two flavors is not zero, we suppose that the violation could be observed in the $\overline{\nu_e}$ survival channel at Daya-Bay. In order to estimate the significance from events number of the violations, we simulated the statistical quantity by creating large sample of pseudodata based on the fitting result of observed $P_{ee}$ values. The pseudodata are generated by a Gaussian distribution model with the means and variances matched to the center values and deviations of the best fit. Each set of simulated data gives an artificial number of LGI violations for $K_3$ and $K_4$, from which we can calculate the level of inconsistency of the predictions between quantum and classical $K_n$.

To estimate the confidence level of these results being inconsistent with a realism expected prediction, we make a fit of histogram filled by predicted LGI violations number under realism model~\ref{eq:3_5} to a beta-binomial distribution, thus to estimate the deviation of classical predictions from the actually observed number of LGI violations. For the actual number of LGI violations (41 in 48 data points), there exists a $6.1\sigma$ deviation from the expected distribution of the classical prediction.

\begin{figure}[htp]
\centering
{
\begin{minipage}[b]{0.55\textwidth}
\includegraphics[width=1\textwidth]{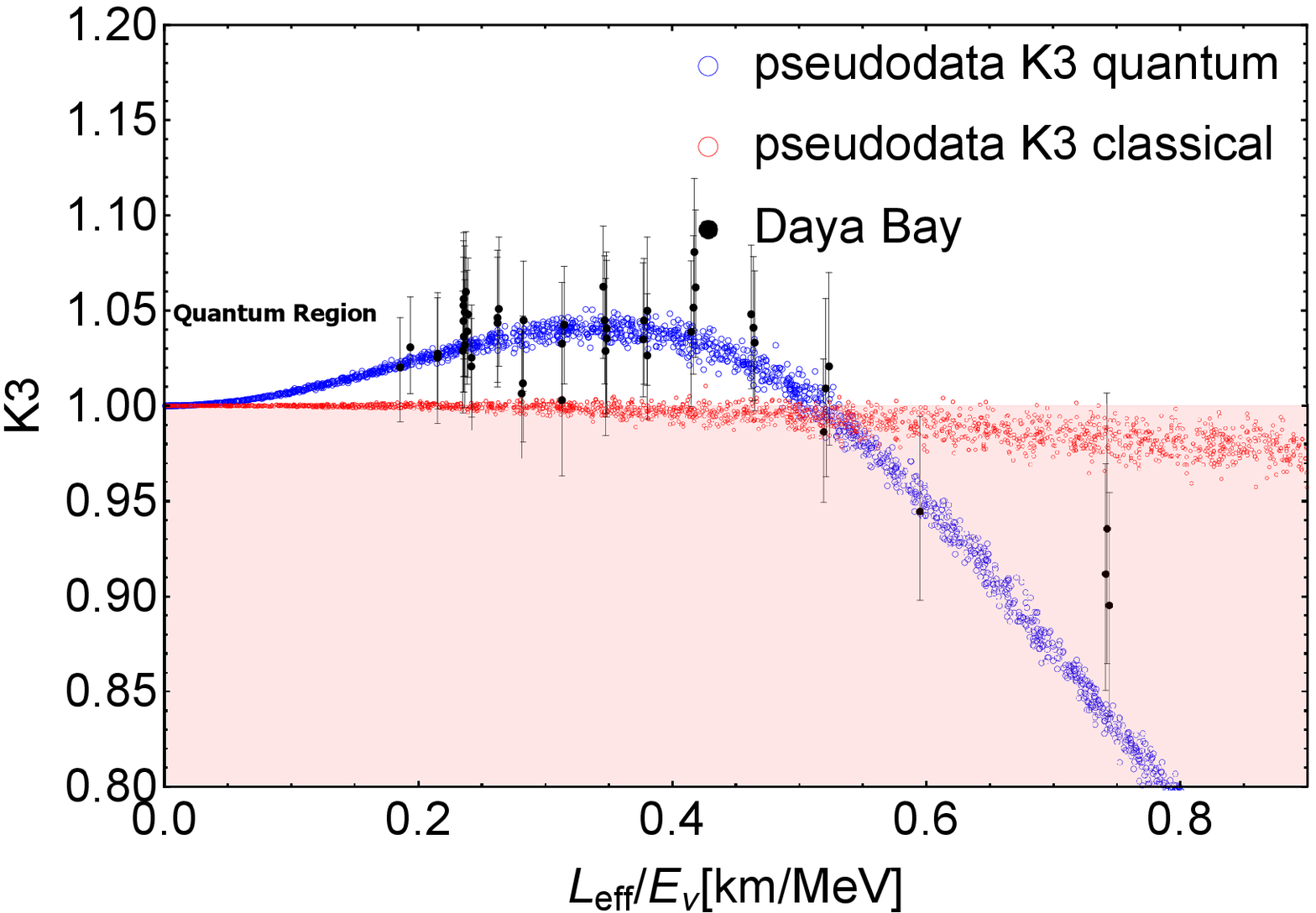}
\end{minipage}
}

{
\begin{minipage}[b]{0.55\textwidth}
\includegraphics[width=1\textwidth]{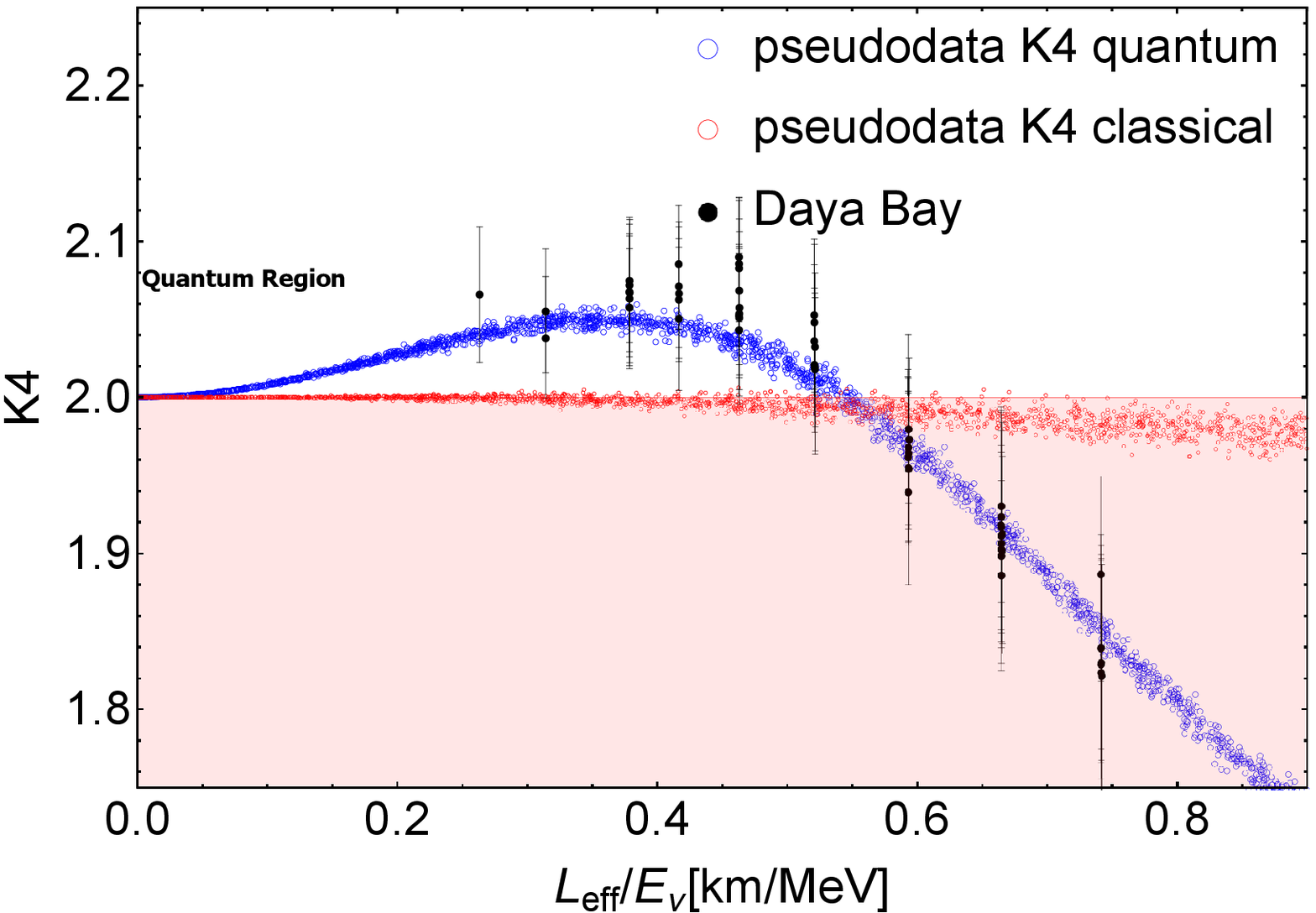}
\end{minipage}
}
\caption{(color online). The distribution of K3 (upper) and K4 (lower) versus the effective propagation length divided by neutrino energy reconstructed from $P_{ee}$. The black dot data show a cluster over the LGI bound. We also show the expected distributions of classical (red circles) and quantum (blue circle) predictions. Note that the $K_3$ and $K_4$ can be multiple values, since there are many triples and quadruple satisfy the phase sum rule.}
\label{Fig:3_3}
\end{figure}

A similar statistical test is made for LGI $K_4$. Using the filter of the phase sum rule described above, we get a number of 30 (in total of 56 data points) exceed the classical limits. As Fig.~\ref{Fig:3_3} shows, there are obvious clusters of points over the classical bound of $K_3$ and $K_4$. The discrepancy between the observed events number and the classical predicted events originating from the fluctuation is very clear. Our $K_4$ data also possesses $6\sigma$ deviation from the classical prediction.   
\begin{figure}[htp]
\centering
{
\begin{minipage}[b]{0.55\textwidth}
\includegraphics[width=1\textwidth]{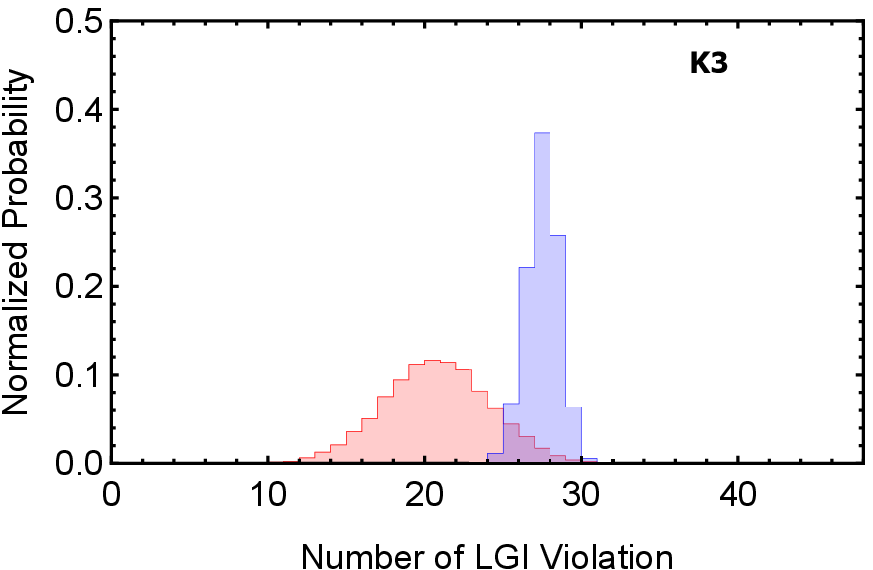}
\end{minipage}
}

{
\begin{minipage}[b]{0.55\textwidth}
\includegraphics[width=1\textwidth]{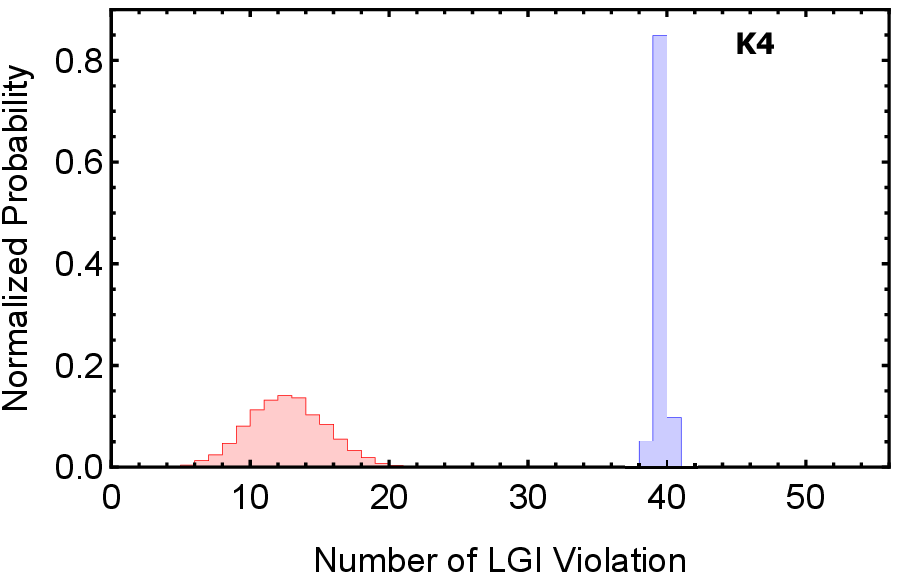}
\end{minipage}
}
\caption{(color online). The histograms of number of K3 (upper) and K4 (lower) values that violate the LGI bound. The left curves with red filling indicate the expected classical distributions, while the right cures with blue filling indicate the quantum corresponding quantity.}
\label{Fig:3_2}
\end{figure}

\section{Discussion}
The results mentioned above clearly constrain the validity of quantum mechanics in such a macroscopic area. Values of LGtI $K_3$ and $K_4$ are violated with the QM's prediction at the confidence level of over $6\sigma$ comparing with the classical bound for the neutrino $\theta_{13}$ mixing in our estimation. We get to present that anti-electron neutrino oscillations also violate the limits of Leggett-Garg inequality. The detected violations act as a new affirmation of quantum nonlocality existing in neutrino system during its long range propagation. These violations were observed over the near and far detectors placed at three experimental halls (EHs) with the baseline long enough to make the test not being a Bell-like inequality test. Besides, it should be worthwhile to make a detailed data analysis on the Daya-Bay experiment involving three-flavor neutrino oscillation, in order to achieve more data points of LGI $K_3$ and $K_4$. It could be worth to test the quantum mechanics in such a weak interaction context. Although tests of incompatible of LGI and QM have been achieved by photonics and electronic experiments~\cite{PhysRevB.86.085418, PhysRevA.87.052115}, nuclear spin qubits~\cite{PhysRevLett.107.090401} and even condensed states~\cite{PhysRevLett.115.113002}, there are few reports of LGI violation in particle physics. Even though the MINOS and Daya-Bay experimental setup show the LGtI violations, these two experiments are all in the context of two flavor neutrino oscillation, which can not reveal the CP violation. Since the entanglement exists between a pair of neutral meson and anti-meson, which will violate the Bell inequality~\cite{Hiesmayr2012}, three-flavor oscillation analysis involving neutrinos and anti-neutrinos may shed light on the study of CP-violating phase.~\cite{Naikoo:2017fos}    

\noindent{\bf Acknowledgments}:
The authors thank Jarah Evslin for helpful discussions and thank J. A. Formaggio for the illuminating suggestions and answers for our questions. This work was supported by the Key Research Program of Frontier Sciences, CAS, under the Grants Number NO. QYZDY-SSW-SLH006 of Chinese Academy of Sciences.

\end{document}